\pdfoutput=1
\documentclass[%
 reprint,
 superscriptaddress,
 preprintnumbers,
 bibnotes,
 amsmath,amssymb,
 aps,
]{revtex4-1}

\usepackage[dvips]{graphicx}
\usepackage{color}
\usepackage{array}
\usepackage{slashed}
\usepackage{tikz}
\usepackage{hyperref}
\usepackage[caption=false]{subfig}
\usepackage{arydshln}
\usepackage{comment}
\usepackage{yhmath}

\usetikzlibrary{shapes.geometric,arrows,decorations.pathmorphing,decorations.markings,patterns}

\usepackage{color}

\newif\ifdraft
\drafttrue

\def\spa#1.#2{\left\langle#1\,#2\right\rangle}
\def\spb#1.#2{\left[#1\,#2\right]}

\font\tenshuffle=shuffle10 \font\sevenshuffle=shuffle7 \font\fiveshuffle=shuffle7 at 5pt
\def\shuffle{{%
\def\Dshuffle{\mathbin{\hbox{\tenshuffle\char'001}}}%
\def\Sshuffle{\mathbin{\hbox{\sevenshuffle\char'001}}}%
\def\SSshuffle{\mathbin{\hbox{\fiveshuffle\char'001}}}%
\mathchoice{\Dshuffle}{\Dshuffle}{\Sshuffle}{\SSshuffle}}}

\newcommand{\eq}{\begin{equation}}
\newcommand{\eqe}{\end{equation}}
\newcommand{\eqa}{\begin{eqnarray}}
\newcommand{\eqae}{\end{eqnarray}}

\newcommand{\bea}{\begin{eqnarray}}
\newcommand{\eea}{\end{eqnarray}}

\newcommand*{\halfway}{0.5*\pgfdecoratedpathlength+.5*3pt}
\newcommand*{\halfwayb}{0.166*\pgfdecoratedpathlength+.5*3pt}
\newcommand*{\halfwayc}{0.833*\pgfdecoratedpathlength+.5*3pt}
\newcommand*{\halfwayd}{0.25*\pgfdecoratedpathlength+.5*3pt}
\newcommand*{\halfwaye}{0.5833*\pgfdecoratedpathlength+.5*3pt}
\newcommand*{\halfwayf}{0.9166*\pgfdecoratedpathlength+.5*3pt}
\newcommand*{\halfwayg}{0.75*\pgfdecoratedpathlength+.5*3pt}

\newbox\charbox
\newbox\slabox
\def\s#1{{      
        \setbox\charbox=\hbox{$#1$}
        \setbox\slabox=\hbox{$/$}
        \dimen\charbox=\ht\slabox
        \advance\dimen\charbox by -\dp\slabox
        \advance\dimen\charbox by -\ht\charbox
        \advance\dimen\charbox by \dp\charbox
        \divide\dimen\charbox by 2
        \raise-\dimen\charbox\hbox to \wd\charbox{\hss/\hss}
        \llap{$#1$}
}}

\def\be{\begin{equation}}
\def\ee{\end{equation}}
\def\ba{\begin{eqnarray}}
\def\ea{\end{eqnarray}}
\def\nl{\nonumber\\}

\def\CP1{\mathbb{CP}^1}
\def\SL2C{\mathrm{SL}(2,\mathbb{C})}

\def\Z2{\mathbb{Z}_2}

\def\su2{{SU(2)}}

\def\[{\left[}
\def\]{\right]}

\def\Cdot{\!\cdot\!}

\def\s{\sigma}

\def\({\left(}
\def\){\right)}
\def\[{\left[}
\def\]{\right]}

\def\<{\langle}
\def\>{\rangle}

\def\i2{\frac{i}{2}}

\def\2F1{\,_2{\rm F}_1}

\definecolor{mygreen}{rgb}{0,0.4,0}

\begin{document}

\title{
String amplitudes from field-theory amplitudes and {\it vice versa}
} 

\author{Song He}
\email{songhe@itp.ac.cn}
\affiliation{CAS Key Laboratory of Theoretical Physics, Institute of Theoretical Physics, Chinese Academy of Sciences, Beijing 100190, China}
\affiliation{School of Physical Sciences, University of Chinese Academy of Sciences, No.19A Yuquan Road, Beijing 100049, China
}
\author{Fei Teng}
\email{fei.teng@physics.uu.se}
\affiliation{Department of Physics and Astronomy, Uppsala University, 75108 Uppsala, Sweden}

\author{Yong Zhang}
\email{yongzhang@itp.ac.cn}
\affiliation{CAS Key Laboratory of Theoretical Physics, Institute of Theoretical Physics, Chinese Academy of Sciences, Beijing 100190, China}
\affiliation{School of Physical Sciences, University of Chinese Academy of Sciences, No.19A Yuquan Road, Beijing 100049, China
}

\preprint{UUITP-58/18}

\begin{abstract}
We present an integration-by-parts reduction of any massless tree-level string correlator to an equivalence class of {\it logarithmic functions}, which can be used to define a field-theory amplitude via a Cachazo-He-Yuan (CHY) formula. The string amplitude is then shown to be the double copy of the field-theory one and a special disk/sphere integral. The construction is generic as it applies to any correlator that is a rational function of correct SL$(2)$ weight. By applying the reduction to open bosonic/heterotic strings, we get a closed-form CHY integrand for the $(DF)^2+\text{YM}+\phi^3$ theory.
\end{abstract}

\maketitle

\section{Introduction}  

\noindent
Recent years have seen enormous advances in understanding structures of scattering amplitudes in quantum field theories ({\it c.f.}~\cite{ArkaniHamed:2012nw,Henn:2014yza,Elvang:2015rqa}), and many crucial insights have originated from string theory. The field-theory limit of the celebrated Kawai-Lewellen-Tye (KLT) relations~\cite{Kawai:1985xq} between tree amplitudes in open and closed string theory gives double copy relations linking gauge-theory amplitudes to gravitational ones~\cite{Bern:1998sv}. Based on a remarkable duality between color and kinematics due to Bern, Carrasco and Johansson (BCJ)~\cite{Bern:2008qj}, double copy has been extended to quantum level and become the state-of-the-art method for multi-loop calculations in supergravities~\cite{Bern:2010ue}. Apart from the original KLT relations, string theory has provided constructions of amplitude representation that respects color-kinematics duality at tree and loop level~\cite{Mafra:2011kj,Mafra:2011nv,Mafra:2011nw,Mafra:2014gja, He:2015wgf}, and amplitude relations in gauge theory, {\it i.e.} BCJ relations~\cite{BjerrumBohr:2009rd,Stieberger:2009hq}, and 
beyond~\cite{Stieberger:2016lng,Schlotterer:2016cxa}.

More surprisingly, it has been realized that tree-level open- and closed-superstring amplitudes themselves can be obtained via a double copy~\cite{Broedel:2013tta}. By rearranging the worldsheet correlator with massless states in type-I theory~\cite{Mafra:2011nw,Mafra:2011nv}, we can decompose disk amplitudes into field-theory KLT products of super-Yang-Mills (SYM) amplitudes and universal basis of disk integrals called $Z$ integrals. The former contain the full polarization information but no nontrivial $\alpha'$-dependence, while the latter encode $\alpha'$ expansion and are interpreted as amplitudes in an effective field theory dubbed $Z$ theory~\cite{Carrasco:2016ldy,Mafra:2016mcc,Carrasco:2016ygv}. Very recently, this has been generalized to bosonic and heterotic strings~\cite{Huang:2016tag,Azevedo:2018dgo}: now the field-theory amplitudes also contain tachyon poles, and they were conjectured to come from the $(DF)^2 + {\rm YM} + \phi^3$ Lagrangian, with $\alpha'$ related to its mass parameter~\cite{Johansson:2017srf}. With $Z$ integrals replaced by certain sphere integrals, we can get closed-string amplitudes via a double copy from these field theories~\cite{Schlotterer:2012ny}. 

An alternative framework that manifests and extends the double copy in field theories is the Cachazo-He-Yuan (CHY) formulation~\cite{Cachazo:2013hca, Cachazo:2013iea}, which expresses tree amplitudes in a large class of massless theories based on scattering equations~\cite{Cachazo:2013gna}. Together with loop-level generalizations~\cite{Adamo:2013tsa,Geyer:2015bja,Cachazo:2015aol,Geyer:2016wjx,Geyer:2018xwu}, these have led to new double-copy realization of various theories~\cite{Cachazo:2014xea%
}, and one-loop extensions of KLT and amplitude relations~\cite{He:2016mzd,He:2017spx}. It is now well established that tree and loop CHY formulas can be derived from worldsheet models known as ambitwistor string theory~\cite{Mason:2013sva,Casali:2015vta}, where CHY integrand is given by correlators therein. Despite significant progress~\cite{Siegel:2015axg,Casali:2016atr,Azevedo:2017yjy,Mizera:2017rqa,Bjerrum-Bohr:2014qwa}, it is fair to say that the origin of ambitwistor strings, and the relations to strings at finite $\alpha'$ remain mysterious. A related outstanding question is how such double copies for strings can be understood via ambitwistor strings, especially for bosonic and heterotic strings which would require CHY formulas with nontrivial $\alpha'$ dependence.

In this letter, we present a double-copy construction for any massless string amplitude in terms of field-theory amplitudes defined by a CHY formula, where the integrand can be directly obtained from the original string correlator. We show algorithmically how to reduce a completely general rational function with correct SL$(2)$ weight (which we refer to as a string correlator), via integration by parts (IBP) to an equivalent class of  logarithmic functions, and any of them serves as a CHY half-integrand that gives desired field-theory amplitudes. Once the logarithmic class is reached, the next step is to use scattering equations (SE) to obtain equivalent half-integrands, which are no longer logarithmic but usually take a more compact form and make some properties more manifest. The same holds for closed string correlators where we reduce the holomorphic part by IBP.

As the main application illustrating the power of this method, we perform IBP for all scalar-gluon correlators of compactified open bosonic strings, or equivalently the holomorphic part in heterotic strings, to produce logarithmic functions as CHY half-integrands. By using SE, we further present a remarkably simple formula for the half-integrand to any multiplicities which gives amplitudes for the single-trace sector of the $(DF)^2\,+\,{\rm YM}\,+\,\phi^3$ theory. The formula for multitrace amplitudes and pure gluon ones are contained in our result as well, thus it gives CHY representations for all tree amplitudes in this theory. 

\section{CHY integrand from string correlator}

The double copy construction at tree-level can be most conveniently expressed by KLT product of color-ordered amplitudes: ${\cal M}_C={\cal M}_A \otimes {\cal M}_B$ is defined as 
\be
{\cal M}_C=\sum_{\alpha,\beta\in S_{n{-}3}} {\cal M}_A (\alpha)~S[\alpha|\beta]~{\cal M}_B (\beta)\,,
\ee
where the $(n{-}3)!$ dimensional matrix $S[\alpha|\beta]$ is the so-called momentum kernel~\cite{BjerrumBohr:2010yc} for orderings $\alpha, \beta$ in a minimal basis. Instead of writing $S$ explicitly, it suffices to say that it is the inverse of the matrix formed by double-partial amplitudes in bi-adjoint $\phi^3$ theory: $S=m^{-1}$~\cite{Cachazo:2013iea}, for any choice of basis such that $m$ is invertible.  

A generic massless open-string tree amplitude is given by a disk integral with $\rho \in S_n/Z_n$ denoting the ordering for integration domain $z_{\rho(i)}<z_{\rho(i{+}1)}$:
\be\label{string}
{\cal M}_n^{\rm string}(\rho)=\int_{\rho}\underbrace{\frac{d^n z}{{\rm vol}\,\text{SL}(2, \mathbb{R})} \overbrace{\prod_{i<j} |z_{i \,j}|^{s_{ij}}}^{:=\text{KN}}}_{:= d\mu_n^{\rm string}}
~{\cal I}_n(z)\,,
\ee
where $s_{ij}:= \alpha' k_i \Cdot k_j$, $z_{ij}:= z_i-z_j$; one can fix three punctures, {\it e.g.} $(z_1, z_{n-1}, z_n)=(0, 1, \infty)$, using SL$(2, \mathbb{R})$ redundancy, and the product in the Koba-Nielsen factor is over $1\leqslant i<j\leqslant n{-}1$ with this fixing. After stripping it off, the (reduced) {\it string correlator} ${\cal I}_n$ is a rational function of $z$'s which depends on details of vertex operators. The only requirement here is that it has the correct SL$(2)$ weight: with $z_a \to \frac{\alpha +\beta z_a}{\gamma+\delta z_a}$ and $\alpha\delta - \beta\gamma = 1$, it must transform as ${\cal I}_n \to \prod_{a=1}^n (\gamma+\delta z_a)^2 {\cal I}_n$. 

\subsection{General claim}

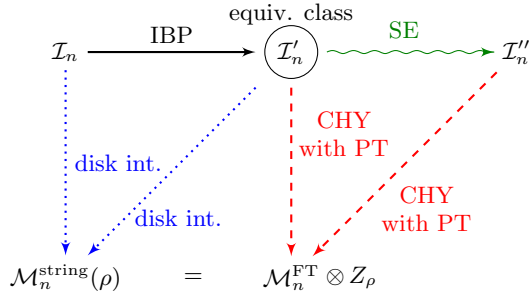
\begin{figure}[!htb]
\tikzstyle{line} = [draw, thick, -latex']

 \begin{tikzpicture}[node distance = 3cm, auto,decoration={coil}]
  \node at (-1,0)  (in) {${\cal I}_n$
    };
    \node at (2,0) (inp) {  \tikz{\draw(0,0) circle (10pt);  \node at (0,0) {${\cal I}_n'$};}};
      \node at (5,0) (inpp) {${\cal I}_n''$
    };
    \node at (2,0.5) {equiv. class};
    
     \node at (-1,-3) (mst) {${\cal M}_n^{\rm string}(\rho)$
    };
      \node at (2,-3) (mft) {${\cal M}_{n}^{\rm FT}$
    };
    
       \node  at  (2.8  ,-3)  {$\otimes\, Z_\rho$
    };

     \node at (0.5,-2.2) {\color{blue} disk int.} ;
       
    \node at (0.7,-3) {=
    };
    
    \path [line] (in) -- node {IBP} (inp);
      
     
       \path [line,dotted,blue] (in) -- node{disk int.} (mst);
       
         \path [line,dotted,blue] (inp) -- (mst);
    
      \draw[decorate, green!50!black,decoration={aspect=0.2, segment length=2.8mm, amplitude=.2mm}] (inp) --(inpp);
        \path [line,green!50!black] (4.6,0) --  (4.7,0);
      
      \node at (3.5,.3) {\color{green!50!black} {SE}};
      
       \path [line,dashed,red] (inp) --(mft);
       
       \node at (2.7,-.9) {\color{red}CHY};
       \node at (2.7,-1.27){\color{red} with PT};

        \path [line,red,dashed] (inpp) --(mft);
          \node at (3.8,-1.9) {\color{red}CHY};
       \node at (3.8,-2.27){\color{red} with PT};
      
      \end{tikzpicture}
      \caption{Double copy for string amplitudes, and CHY half-integrands from string correlator via IBP and SE. 
       \label{inin}}
\end{figure}

Our general claim, as summarized in Fig.~\ref{inin}, is that ${\cal M}_n^{\rm string}(\rho)$ is the double copy of $Z$ integrals (disk integral with a Parke-Taylor (PT) factor~\cite{Broedel:2013tta}) 
\be\label{ztheory}
Z_\rho(\pi)=\!\int_\rho \frac{d\mu_n^{\rm string}}{z_{\pi_1\pi_2}z_{\pi_2\pi_3}\cdots z_{\pi_n\pi_1}}:=\!\int_\rho d\mu_n^{\rm string}\,{\rm PT}(\pi)
\ee
with color-ordered amplitudes in a field theory. The latter is defined by a CHY formula, with integrands given by a PT factor as defined above, and a logarithmic function ${\cal I}_n'$ obtained from an IBP reduction of ${\cal I}_n$:
\be\label{field}
{\cal M}_n^{\rm FT} (\rho):=\!\int\!\!\underbrace{\frac{d^n z}{{\rm vol}\,\text{SL}(2, \mathbb{C})}\prod_i{}' \delta(\sum_{j\neq i} \frac{s_{i\,j}}{z_{i\,j}})}_{:=d\mu_n^{\rm CHY}}{\rm PT}(\rho)\,{\cal I}'_n(z)\,.
\ee 
Here the integrals are localized by the $n{-}3$ delta functions imposing scattering equations~\cite{Cachazo:2013gna,Cachazo:2013hca}. Logarithmic functions are defined to have only logarithmic singularities on boundaries of the moduli space of $n$-punctured Riemann spheres;
equivalently, such a function can be written as a linear combination of PT factors~\cite{Mizera:2017cqs,Arkani-Hamed:2017tmz,Arkani-Hamed:2017mur}.

The idea that string amplitude can be written as a double copy has been realized in various theories~\cite{Mafra:2011nv,Mafra:2011nw,Huang:2016tag,Azevedo:2018dgo} and expected to hold for generic $\mathcal{I}_n$. For example, by using intersection theory, we can obtain $\mathcal{I}'_n$ from $\mathcal{I}_n$ by computing intersection numbers~\footnote{See Sebastian Mizera's talk at \href{https://indico.cern.ch/event/646820/contributions/2992864/attachments/1671110/2680821/Intersections.pdf}{Amplitude 2018}}. In this work, we will give a concrete and streamlined IBP algorithm that reduces a generic $\mathcal{I}_n$ explicitly to a class of logarithmic functions; the algorithm lands ${\cal I}_n'$ on a nonminimal basis of ${\rm PT}$ factors, making further simplifications accessible. 

We actually have an equivalence class of logarithmic functions denoted as ${\cal I}'_n \overset{\rm IBP}\cong {\cal I}_n$: any ${\cal I}_n'$ gives the same string integral as that of ${\cal I}_n$. As we will show shortly, ${\cal I}_n'$ gives exactly ${\cal M}_n^{\rm FT}$ needed for the double copy, via CHY formula with a PT factor; any ${\cal I}_n'$ in the class gives the same result since they are also in an equivalent class on the support of SE. While the explicit form of ${\cal I}_n'$ gets more complicated as $n$ grows, we emphasize that it can usually be simplified greatly by use of SE (see Fig.~\ref{inin}). The manipulation from ${\cal I}_n'$ to ${\cal I}_n''$ is in a sense the reversed one as going from ${\cal I}_n$ to ${\cal I}_n'$: while being non-logarithmic, usually $\mathcal{I}''_n$ allows an all-multiplicity expression that makes other properties, such as gauge invariance, manifest. The primary example of this double copy is ``type-I $\,=\,Z~\otimes$ SYM"~\cite{Mafra:2011nw,Mafra:2011nv}, where the class of logarithmic functions from type-I correlator can be used to produce SYM amplitudes. It is now well known that the CHY half-integrand for external gluons, which takes a remarkably simple form ${\cal I}_n''={\rm Pf}' \pmb{\Psi}_n $~\footnote{We refer the readers to the original paper~\cite{Cachazo:2013hca} for the definition of the $2n\times 2n$ matrix $\pmb{\Psi}_n$ and more details on CHY formulas.}, is equivalent to ${\cal I}_n'$ on the support of SE.

Last but not least, one can generalize the double copy to closed-string amplitudes by replacing $Z$ integrals by certain sphere integrals. As conjectured in~\cite{Schlotterer:2012ny, Stieberger:2013wea, Stieberger:2014hba} and proven in \cite{Schlotterer:2018zce,Brown:2018omk}, the latter can be obtained as the single-valued (sv) projection
~\cite{Schnetz:2013hqa, Brown:2013gia} of open-string amplitudes, and {\it e.g.,} we have ``type-II $=$ sv(type-I) $\otimes$ SYM". All of our discussions hold for such double copies where IBP is applied to the holomorphic part of closed-string correlator ${\cal I}_n (z)$ (with the antiholomorphic part untouched): schematically we write~\cite{Azevedo:2018dgo} ${\cal M}_n^{\rm closed}={\cal M}_n^{\rm FT} \otimes {\rm sv} ({\cal M}_n^{\rm open})$, with the CHY half-integrand for ${\cal M}_n^{\rm FT}$ obtained from the holomorphic correlator, ${\cal I}_n' \overset{\rm IBP}{\cong} {\cal I}_n$.  

\begin{figure}[tb]
    \subfloat[tadpole]{\begin{tikzpicture}[baseline={([yshift=-0.ex]current bounding box.center)},every node/.style={font=\footnotesize}]
    \draw [decoration={markings, mark=at position \halfwayb with {\arrow{>}}, mark=at position \halfway with {\arrow{>}},mark=at position \halfwayc with {\arrow{>}}},postaction={decorate}] (0,0) circle (0.5);
    \draw (0.5,0) -- (1,0);
    \filldraw (120:0.5) circle (1pt) node[below right=-1.5pt]{$1$} (-120:0.5) circle (1pt) node[above right=-1.5pt]{$2$} (0.5,0) circle (1pt) node[left=0pt]{$3$} (1,0) circle (1pt) node[below left=-1.5pt]{$4$};
    \path [decoration={markings, mark=at position \halfway with {\arrow{>}}},postaction={decorate}] (1,0) -- (0.5,0);
    \draw [decoration={markings, mark=at position \halfway with {\arrow{>}}},postaction={decorate}] (1.5,0) -- (1,0);
    \filldraw (1.5,0) circle (1pt) node[below=0pt]{$5$} (0.5,-0.5) circle (1pt) node[right=0pt]{$6$};
    \node at (-120:1) [left=0pt]{$\phantom{7}$};
    \node at (120:1) [left=0pt]{$\phantom{7}$};
    \node at (0,1) [left=0pt]{$\phantom{8}$};
    \node at (0.5,-1) {$(z_7\rightarrow\infty)$};
    \end{tikzpicture}\label{tadpole}}\quad
    \subfloat[multibranch]{\begin{tikzpicture}[baseline={([yshift=-0.ex]current bounding box.center)},every node/.style={font=\footnotesize}]
    \draw [decoration={markings, mark=at position \halfwayd with {\arrow{>}}, mark=at position \halfwaye with {\arrow{>}},mark=at position \halfwayf with {\arrow{>}}},postaction={decorate}] (0,0) circle (0.5);
    \filldraw (150:0.5) circle (1pt) node[below right=-1.5pt]{$3$} -- (150:1) circle (1pt) node[above=0pt]{$5$} (30:0.5) circle (1pt) node[below left=-1.5pt]{$1$} -- (30:1) circle (1pt) node[above=0pt]{$4$} (0,-0.5) circle (1pt) node[above=0pt]{$2$} (0,-0.75) circle (1pt) node[left=0pt]{$6$};
    \path [decoration={markings, mark=at position \halfway with {\arrow{>}}},postaction={decorate}] (150:1) -- (150:0.5);
    \path [decoration={markings, mark=at position \halfway with {\arrow{>}}},postaction={decorate}] (30:1) -- (30:0.5);
    \node at (1,0) {};
    \node at (-1,0) {};
    \node at (0,1) [left=0pt]{$\phantom{8}$};
    \node at (-120:1) [left=0pt]{$\phantom{7}$};
    \node at (0,-1) {$(z_7\rightarrow\infty)$};
    \end{tikzpicture}\label{multib}}\quad
    \subfloat[generic]{\begin{tikzpicture}[baseline={([yshift=-0.ex]current bounding box.center)},every node/.style={font=\footnotesize}]
    \draw [decoration={markings, mark=at position \halfwayb with {\arrow{>}}, mark=at position \halfway with {\arrow{>}},mark=at position \halfwayc with {\arrow{>}}},postaction={decorate}] (0,0) circle (0.5);
    \draw (120:0.5) -- ++(120:0.5) (-120:0.5) -- ++(-120:0.5);
    \filldraw (120:0.5) circle (1pt) node[below right=-1.5pt]{$1$} (0:0.5) circle (1pt) node[left=0pt]{$3$} (-120:0.5) circle (1pt) node[above right=-1.5pt]{$2$} (120:1) circle (1pt) node[left=0pt]{$6$} (-120:1) circle (1pt) node[left=0pt]{$7$};
    \draw [decoration={markings, mark=at position \halfway with {\arrow{>}},mark=at position 1 with {\arrow{>}}},postaction={decorate}] (1.25,0) circle (0.5);
    \filldraw (1.25,-0.5) circle (1pt) node[above=0pt]{$5$} (1.25,0.5) circle (1pt) node[below=0pt]{$4$} -- ++(0,0.5) circle (1pt) node[left=0pt]{$8$};
    \filldraw (2,-0.5) circle (1pt) node[below=0pt]{$10$} -- (2,0.5) circle (1pt) node[above=0pt]{$9$};
    \path [decoration={markings, mark=at position \halfway with {\arrow{>}}},postaction={decorate}] (2,0.5) -- (2,-0.5);
    \path [decoration={markings, mark=at position \halfway with {\arrow{>}}},postaction={decorate}] (120:1) -- (120:0.5);
    \path [decoration={markings, mark=at position \halfway with {\arrow{>}}},postaction={decorate}] (-120:1) -- (-120:0.5);
    \path [decoration={markings, mark=at position \halfway with {\arrow{>}}},postaction={decorate}] (1.25,1) -- (1.25,0.5);
    \node at (0.675,-1) {$(z_{11}\rightarrow\infty)$};
    \end{tikzpicture}\label{generic}}
    \caption{Examples of rational functions of $z_{ij}$ with no numerators or connected subcycles after gauge fixing. For this case, gauge fixing always results in a single labeled tree.}
\end{figure}
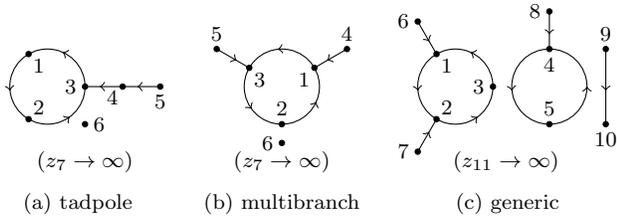

\subsection{Derivation: IBP reduction and double copy}\label{sec:IBPproof}

We can conveniently represent any rational functions involving only $z_{ij}$'s by a graph: we denote each $z_{ij}$ in the denominator (numerator) by a solid (dashed) line directed from node $i$ to $j$. With the gauge fixing, {\it e.g.} $z_n\rightarrow\infty$, all the labeled trees are logarithmic functions since we can write them as a sum of PT factors~\cite{Gao:2017dek}.

The derivation of our general claim consists of two steps. We first give a method for reducing any string correlator to logarithmic functions via IBP relations. Next, we show that such logarithmic functions are precisely the CHY integrands for those field theories that enter the string-amplitude double copy mentioned above.

\paragraph{IBP reduction} 
We start with a simple example ${\cal I}_4=\frac{1}{z_{12}^2 z_{34}^2}$, which, after we gauge fix $(z_1,z_3,z_4)=(0,1,\infty)$ and subtract the total derivative $\partial_{z_2}\frac{z_2^{s-1}(1-z_2)^t}{s-1}$, reduces to a logarithmic function, in this case a PT factor after restoring the SL$(2)$ covariance:
\begin{align}\label{I4IBP}
 \frac{1}{z_{12}^2z_{34}^2}\overset{\rm IBP}{\cong}\frac{t}{1-s}\frac{1}{z_{12}z_{23}z_{34}z_{41}}=\frac{t}{1- s} {\rm PT}(1234)\,,
\end{align}
where $s=s_{12}$ and $t=s_{23}$. This example represents the simplest scenario where only a single subcycle and a labeled line are left after gauge fixing $z_n \to \infty$. The next simplest case is represented by \emph{tadpole} graphs, as depicted in Fig.~\ref{tadpole}, in which a single labeled line connects to the subcycle. One can reduce them to a sum of labeled trees by recursively using the IBP relation~\cite{Schlotterer:2016cxa}:
\begin{align}\label{IBPrelation}
    \text{PT}(12\cdots m)&\overset{\rm IBP}{\cong}\Bigg(\sum_{\ell=2}^{m}\sum_{j=m+1}^{n-1}\sum_{\rho\in X\shuffle Y^T}\frac{(-1)^{|Y|}}{1-s_{12\cdots m}}\nonumber\\
    &\quad\times\frac{s_{\ell j}}{z_{1\rho_1}z_{\rho_1\rho_2}\cdots z_{\rho_{|\rho|}\ell}z_{\ell j}} \Bigg)\,,
\end{align}
where $X$ and $Y$ are obtained by matching  $(1,2\cdots m)=(1,X,\ell,Y)$. This relation holds when $\text{PT}(12\cdots m)$ is multiplied by a function
independent of $z_2$ to $z_m$. The total derivative of $z_1$ is not used to derive this relation, and we identify it as the node shared by the subcycle and the tail. For example, Fig.~\ref{tadpole} can be processed as:
\begin{widetext}
\begin{align}\tikzset{baseline={([yshift=-0.7ex]current bounding box.center)},every node/.style={font=\footnotesize}}\label{poll}
  \begin{tikzpicture}
    \draw [decoration={markings, mark=at position \halfwayb with {\arrow{>}}, mark=at position \halfway with {\arrow{>}},mark=at position \halfwayc with {\arrow{>}}},postaction={decorate}] (0,0) circle (0.5);
    \draw (0.5,0) -- (1,0);
    \filldraw (120:0.5) circle (1pt) node[below right=-1.5pt]{$1$} (-120:0.5) circle (1pt) node[above right=-1.5pt]{$2$} (0.5,0) circle (1pt) node[left=0pt]{$3$} (1,0) circle (1pt) node[below =-1.5pt]{$4$};
    \path [decoration={markings, mark=at position \halfway with {\arrow{>}}},postaction={decorate}] (1,0) -- (0.5,0);
    \draw [decoration={markings, mark=at position \halfway with {\arrow{>}}},postaction={decorate}] (1.5,0) -- (1,0);
    \filldraw (1.5,0) circle (1pt) node[below=-1.5pt]{$5$} (0.7,-0.5) circle (1pt) node[below=-1.5pt]{$6$};
    \node at (1.5,0) [right=0pt,font=\normalsize]{$\displaystyle\;\;\overset{\rm IBP}{\cong}$};
    \end{tikzpicture}
    \begin{tikzpicture}
        \node at (-1.8,.5) [font=\normalsize]{$~$};
    \begin{scope}[ yshift=0.5cm]
    \filldraw (180:0.6) circle (0pt) node[left=0pt,font=\normalsize]{$\displaystyle\frac{s_{24}}{1-s_{123}}$} ;
    \draw [decoration={markings, mark=at position .25*\halfway with {\arrow{>}}, mark=at position .75*\halfway with {\arrow{>}},mark=at position 1.25*\halfway with {\arrow{>}},mark=at position 1.75*\halfway with {\arrow{>}}},postaction={decorate}] (0,0) circle (0.6);
    \draw (0.6,0) -- (1.1,0);
    \filldraw (90:0.6) circle (1pt) node[below =-1.5pt]{$3$} (180:0.6) circle (1pt) node[ right=-1.5pt]{$1$} (-90:0.6) circle (1pt) node[above=-1.5pt]{$2$} (0:0.6) circle (1pt) node[left=-1.5pt]{$4$}  (1.1,0) circle (1pt) node[below =-1.5pt]{$5$};
    \path [decoration={markings, mark=at position \halfway with {\arrow{>}}},postaction={decorate}] (1.1,0) -- (0.6,0);
       \filldraw (0.7,-0.5) circle (1pt) node[below=-1.5pt]{$6$} ;
    \end{scope}
    \begin{scope}[xshift=4cm,yshift=.5cm]
    \filldraw (180:0.7) circle (0pt) node[left=0pt,font=\normalsize]{$\displaystyle+\;\frac{s_{25}}{1-s_{123}}$} ;
    \draw [decoration={markings, mark=at position .2*\halfway with {\arrow{>}}, mark=at position .6*\halfway with {\arrow{>}},mark=at position \halfway with {\arrow{>}},mark=at position 1.4*\halfway with {\arrow{>}},mark=at position 1.8*\halfway with {\arrow{>}}},postaction={decorate}] (0,0) circle (0.7);
    \filldraw (0:0.7) circle (1pt) node[left =-1.5pt]{$5$} (72:0.7) circle (1pt) node[ below=-1.5pt]{$4$} (144:0.7) circle (1pt) node[right=-.5pt]{$3$} (216:0.7) circle (1pt) node[right=-.5pt]{$1$}  (-72:0.7) circle (1pt) node[ above=-1.5pt]{$2$};
       \filldraw (0.8,-0.5) circle (1pt) node[below=-1.5pt]{$6$} ;
    \end{scope}
    \begin{scope}[xshift=7cm,yshift=0.5cm]
    \draw (0,0)node {\tikz{\draw [fill=black] (1,0) circle (1pt)}} node [below] {6} node [ left=0,font=\normalsize ] {$\displaystyle+\;\frac{s_{26}}{1-s_{123}}$} --++(0.5,0)node {\tikz{\draw [fill=black] (1,0) circle (1pt)}} node [below] {2}--++(0.5,0)node {\tikz{\draw [fill=black] (1,0) circle (1pt)}} node [below] {1}  --++(0.5,0)node {\tikz{\draw [fill=black] (1,0) circle (1pt)}} node [below] {3}  --++(0.5,0)node {\tikz{\draw [fill=black] (1,0) circle (1pt)}} node [below] {4}  --++(0.5,0)node {\tikz{\draw [fill=black] (1,0) circle (1pt)}} node [below] {5};
    \path [decoration={markings, mark=at position \halfway with {\arrow{>}}},postaction={decorate}] (0.5,0)--(0,0);
    \path [decoration={markings, mark=at position \halfway with {\arrow{>}}},postaction={decorate}] (1,0)--(0.5,0);         \path [decoration={markings, mark=at position \halfway with {\arrow{>}}},postaction={decorate}] (1.5,0)--(1,0);
    \path [decoration={markings, mark=at position \halfway with {\arrow{>}}},postaction={decorate}] (2,0)--(1.5,0);
    \path [decoration={markings, mark=at position \halfway with {\arrow{>}}},postaction={decorate}] (2.5,0)--(2,0);
    \end{scope}
    \begin{scope}[ yshift=-1cm]
    \filldraw (180:0.6) circle (0pt) node[left=0pt,font=\normalsize]{$\displaystyle -\;\frac{s_{14}}{1-s_{123}}$} ;
    \draw [decoration={markings, mark=at position .25*\halfway with {\arrow{>}}, mark=at position .75*\halfway with {\arrow{>}},mark=at position 1.25*\halfway with {\arrow{>}},mark=at position 1.75*\halfway with {\arrow{>}}},postaction={decorate}] (0,0) circle (0.6);
    \draw (0.6,0) -- (1.1,0);
    \filldraw (90:0.6) circle (1pt) node[below =-1.5pt]{$3$} (180:0.6) circle (1pt) node[ right=-1.5pt]{$2$} (-90:0.6) circle (1pt) node[above=-1.5pt]{$1$} (0:0.6) circle (1pt) node[left=-1.5pt]{$4$}  (1.1,0) circle (1pt) node[below =-1.5pt]{$5$};
    \path [decoration={markings, mark=at position \halfway with {\arrow{>}}},postaction={decorate}] (1.1,0) -- (0.6,0);
       \filldraw (0.7,-0.5) circle (1pt) node[below=-1.5pt]{$6$} ;
    \end{scope}
    \begin{scope}[xshift=4cm,yshift=-1cm]
    \filldraw (180:0.7) circle (0pt) node[left=0pt,font=\normalsize]{$\displaystyle-\;\frac{s_{15}}{1-s_{123}}$} ;
    \draw [decoration={markings, mark=at position .2*\halfway with {\arrow{>}}, mark=at position .6*\halfway with {\arrow{>}},mark=at position \halfway with {\arrow{>}},mark=at position 1.4*\halfway with {\arrow{>}},mark=at position 1.8*\halfway with {\arrow{>}}},postaction={decorate}] (0,0) circle (0.7);
    \filldraw (0:0.7) circle (1pt) node[left =-1.5pt]{$5$} (72:0.7) circle (1pt) node[ below=-1.5pt]{$4$} (144:0.7) circle (1pt) node[right=-.5pt]{$3$} (216:0.7) circle (1pt) node[right=-.5pt]{$2$}  (-72:0.7) circle (1pt) node[ above=-1.5pt]{$1$};
       \filldraw (0.8,-0.5) circle (1pt) node[below=-1.5pt]{$6$} ;
    \end{scope}
     \begin{scope}[xshift=7cm,yshift=-1cm]
   \draw (0,0)node {\tikz{\draw [fill=black] (1,0) circle (1pt)}} node [below] {6} node [ left=0,font=\normalsize ] {$\displaystyle-\;\frac{s_{16}}{1-s_{123}}$} --++(0.5,0)node {\tikz{\draw [fill=black] (1,0) circle (1pt)}} node [below] {1}--++(0.5,0)node {\tikz{\draw [fill=black] (1,0) circle (1pt)}} node [below] {2}  --++(0.5,0)node {\tikz{\draw [fill=black] (1,0) circle (1pt)}} node [below] {3}  --++(0.5,0)node {\tikz{\draw [fill=black] (1,0) circle (1pt)}} node [below] {4}  --++(0.5,0)node {\tikz{\draw [fill=black] (1,0) circle (1pt)}} node [below] {5};
     \path [decoration={markings, mark=at position \halfway with {\arrow{>}}},postaction={decorate}] (0.5,0)--(0,0);
         \path [decoration={markings, mark=at position \halfway with {\arrow{>}}},postaction={decorate}] (1,0)--(0.5,0);         \path [decoration={markings, mark=at position \halfway with {\arrow{>}}},postaction={decorate}] (1.5,0)--(1,0);
                     \path [decoration={markings, mark=at position \halfway with {\arrow{>}}},postaction={decorate}] (2,0)--(1.5,0);
                 \path [decoration={markings, mark=at position \halfway with {\arrow{>}}},postaction={decorate}] (2.5,0)--(2,0);
    \end{scope}
    \end{tikzpicture}~.
\end{align}
\end{widetext}
Notice that the tail always gets shortened, such that we can turn tadpole graphs into a sum of labeled trees by repeating the IBP relation~\eqref{IBPrelation}.

By connecting more labeled lines to a tadpole, we get \emph{multibranch} graphs,
as shown in Fig.~\ref{multib}. If we remove a subgraph including part of the subcycle, the rest of the graph turns into a labeled tree, which can be written as a sum of labeled lines through algebraic operations~\cite{Gao:2017dek}. For example, we can view the chain $\{4,1,2,3,5\}$ in Fig.~\ref{multib} as a labeled tree rooted on $1$. Then partial fraction identities lead to
\begin{align}
    \tikzset{baseline={([yshift=-0.7ex]current bounding box.center)},every node/.style={font=\footnotesize}}
    \begin{tikzpicture}
    \draw [decoration={markings, mark=at position \halfway with {\arrow{>}}},postaction={decorate}] (120:1) -- (120:0.5); \draw [decoration={markings, mark=at position \halfway with {\arrow{>}}},postaction={decorate}] (-120:1) -- (-120:0.5);
    \draw [decoration={markings, mark=at position \halfwayd with {\arrow{>}},mark=at position \halfwayg with {\arrow{>}}},postaction={decorate}] (-120:0.5) arc (-120:120:0.5);
    \draw [decoration={markings, mark=at position \halfway with {\arrow{>}}},postaction={decorate}] (120:0.5) ++(-0.3,0) arc (120:240:0.5);
    \filldraw (120:0.5) circle (1pt) node[below=0pt]{$1$} (-120:0.5) circle (1pt) node[above=0pt]{$3$} (0.5,0) circle (1pt) node[left=0pt]{$2$} (120:1) circle (1pt) node[left=0pt]{$4$} (-120:1) circle (1pt) node[left=0pt]{$5$};
    \filldraw (120:0.5) ++(-0.3,0) circle (1pt) node[below=0pt]{$1$} (-120:0.5) ++(-0.3,0) circle (1pt) node[above=0pt]{$3$};
    \end{tikzpicture}=\begin{tikzpicture}
    \begin{scope}[yshift=0.5cm]
    \draw (0,0) -- (2,0);
    \draw [decoration={markings, mark=at position \halfway with {\arrow{>}}},postaction={decorate}] (0,0) .. controls (0.5,0.5) and (1,0.5) .. (1.5,0);
    \foreach \x in {0.125,0.375,0.625,0.875} {
    \path [decoration={markings, mark=at position \x*\pgfdecoratedpathlength+.5*3pt with {\arrow{>}}},postaction={decorate}] (2,0) -- (0,0);
    }
    \filldraw (0,0) circle (1pt) node[below=0pt]{$1$} (0.5,0) circle (1pt) node[below=0pt]{$4$} (1,0) circle (1pt) node[below=0pt]{$2$} (1.5,0) circle (1pt) node[below=0pt]{$3$} (2,0) circle (1pt) node[below=0pt]{$5$};
    \end{scope}
    \begin{scope}[xshift=2.6cm,yshift=0.5cm]
    \draw (0,0) -- (2,0);
    \draw [decoration={markings, mark=at position \halfway with {\arrow{>}}},postaction={decorate}] (0,0) .. controls (0.5,0.5) and (1,0.5) .. (1.5,0);
    \foreach \x in {0.125,0.375,0.625,0.875} {
    \path [decoration={markings, mark=at position \x*\pgfdecoratedpathlength+.5*3pt with {\arrow{>}}},postaction={decorate}] (2,0) -- (0,0);
    }
    \filldraw (0,0) circle (1pt) node[below=0pt]{$1$} (0.5,0) circle (1pt) node[below=0pt]{$2$} (1,0) circle (1pt) node[below=0pt]{$4$} (1.5,0) circle (1pt) node[below=0pt]{$3$} (2,0) circle (1pt) node[below=0pt]{$5$};
    \end{scope}
    \node at (2.3,0.5) [font=\normalsize]{$+$};
    \begin{scope}[yshift=-0.5cm]
    \draw (0,0) -- (2,0);
    \draw [decoration={markings, mark=at position \halfway with {\arrow{>}}},postaction={decorate}] (0,0) .. controls (0.25,0.5) and (0.75,0.5) .. (1,0);
        \foreach \x in {0.125,0.375,0.625,0.875} {
    \path [decoration={markings, mark=at position \x*\pgfdecoratedpathlength+.5*3pt with {\arrow{>}}},postaction={decorate}] (2,0) -- (0,0);
    }
    \filldraw (0,0) circle (1pt) node[below=0pt]{$1$} (0.5,0) circle (1pt) node[below=0pt]{$2$} (1,0) circle (1pt) node[below=0pt]{$3$} (1.5,0) circle (1pt) node[below=0pt]{$4$} (2,0) circle (1pt) node[below=0pt]{$5$};
    \end{scope}
    \begin{scope}[xshift=2.6cm,yshift=-0.5cm]
    \draw (0,0) -- (2,0);
    \draw [decoration={markings, mark=at position \halfway with {\arrow{>}}},postaction={decorate}] (0,0) .. controls (0.25,0.5) and (0.75,0.5) .. (1,0);
        \foreach \x in {0.125,0.375,0.625,0.875} {
    \path [decoration={markings, mark=at position \x*\pgfdecoratedpathlength+.5*3pt with {\arrow{>}}},postaction={decorate}] (2,0) -- (0,0);
    }
    \filldraw (0,0) circle (1pt) node[below=0pt]{$1$} (0.5,0) circle (1pt) node[below=0pt]{$2$} (1,0) circle (1pt) node[below=0pt]{$3$} (1.5,0) circle (1pt) node[below=0pt]{$5$} (2,0) circle (1pt) node[below=0pt]{$4$};
    \end{scope}
    \node at (2.3,-0.5) [font=\normalsize]{$+$};
    \node at (-0.3,-0.5) [font=\normalsize]{$+$};
    \end{tikzpicture}~.
\end{align}
In this way, we can turn all the multibranch graphs into tadpoles, and consequently labeled trees via IBP.

The most generic graph with no $z_{ij}$ in the numerator or connected subcycles has the form of Fig.~\ref{generic}, consisting of a product of tadpole and multibranch graphs, and a single labeled tree~\footnote{If there are more than one labeled trees, there must also be connected  subcycles as a result of the SL$(2)$ weight constraint.}.
Following the above prescription, we can turn a tadpole or multibranch subgraph into labeled trees planted on the other connected components, reducing the number of subcycles by one. This algorithm allows us to recursively eliminate all the subcycles, and the final result contains only labeled trees, thus the function ${\cal I}_n'$ we obtain is logarithmic.  

As we will see shortly, the above discussion applies to open bosonic and heterotic strings. We defer the treatment of generic rational functions involving numerators and connected subcycles to Appendix~\ref{appa}.

\paragraph{String double copy} Now we prove that $\mathcal{M}^{\rm FT}_n$ given by the $\mathcal{I}'_n$ obtained above is indeed the one in the double copy ${\cal M}_n^{\rm string}={\cal M}_n^{\rm FT} \otimes Z_n$. As a logarithmic function, we can write ${\cal I}'_n=\sum_\pi N_\pi {\rm PT}(\pi)$, but the coefficients $N_\pi$ are not unique since the PT factors are not linearly independent. As shown for open strings~\cite{Mafra:2011nv,Broedel:2013tta} and CHY integrals~\cite{Cachazo:2013iea}, they satisfy BCJ relations by IBP and SE equivalence moves respectively. Consequently, any of them can be expressed in terms of the $(n{-}3)!$ dimensional minimal basis:
\begin{align*}
{\rm PT}(\pi)\overset{\text{IBP/SE}}{\cong}\sum_{\beta \in S_{n{-}3}} \Bigg[\sum_{\alpha \in S_{n{-}3}} m(\pi|\alpha) S[\alpha|\beta]\Bigg]{\rm PT}(\beta)\,.
\end{align*}
By using Eq.~\eqref{string} and~\eqref{ztheory}, we have
\begin{align*}
{\cal M}_n^{\rm string}(\rho)=
\sum_{\alpha, \beta \in S_{n{-}3}}\Bigg[\sum_\pi N_\pi\,m(\pi | \alpha) \Bigg]S[\alpha|\beta]\,Z_\rho(\beta)
\end{align*}
for any $\rho\in S_n/Z_n$. The sum in the bracket is a field-theory amplitude defined by CHY formula: recalling that $m(\pi |\alpha)=\!\int\! d\mu_n^{\rm CHY} {\rm PT}(\pi) {\rm PT}(\alpha)$, thus we have
\be\label{FT}
\sum_\pi N_\pi\,m(\pi |\alpha)=\int d\mu_n^{\rm CHY} \underbrace{\sum_\pi N_\pi\,{\rm PT}(\pi)}_{{\cal I}_n'} {\rm PT}(\alpha)\,,
\ee
which gives exactly ${\cal M}_n^{\rm FT}(\alpha)$ as in Eq.~\eqref{field}. 
Note that Eq.~\eqref{FT} makes it manifest that ${\cal M}_n^{\rm FT}(\rho)$ satisfies BCJ relations, which is necessary for a consistent KLT double copy. One can also define ${\cal M}_n^{\rm FT}$ by the LHS, which can be rewritten in a form that respects color/kinematics duality with BCJ numerators given by linear combinations of $N_\pi$'s. However, we find it extremely useful to turn to CHY formula on the RHS: from our explicit result ${\cal I}_n'$, we can use SE to simplify it to some  ${\cal I}_n''$, for which it is possible to obtain a closed form.

One can apply the IBP reduction to type-I correlators ${\cal I}_n={\cal K}_n (z)$ and obtain logarithmic functions, where the coefficients $N_\pi$ give the BCJ numerators for SYM~\cite{Mafra:2011nw}. For gluons, the standard CHY half-integrand ${\cal I}_n''={\rm Pf}' {\pmb\Psi}_n$ can also be reduced via SE to ${\cal I}_n'$, giving equivalent BCJ numerators~\cite{Du:2017kpo}. Remarkably, these BCJ numerators are free of any tachyon poles (though they appear in intermediate steps as in above examples), and unlike ${\cal I}_n$, both ${\cal I}_n'$ and ${\cal I}_n''$ are homogenous in $\alpha'$, all of which are of course guaranteed by the worldsheet supersymmetry. Next we will turn to bosonic/heterotic strings, where BCJ numerators and ${\cal I}_n'$ contains tachyon poles and nontrivial $\alpha'$ dependence. We will see that our two-step method again gives a remarkably simple formula for ${\cal I}''_n$. 

\section{New formulas for bosonic and heterotic strings}\label{sec:heterotic}

Let's first review correlators for (compactified) open bosonic strings and heterotic strings. Since the former is identical to the holomorphic part of the latter, we will focus on the heterotic-string case and everything carries over to the open-string case. By a slight abuse of notation, we denote closed-string measure as $| d\mu_n^{\rm string} |^2:=\frac{d^{n} zd^n\bar{z}~ {\rm KN}}{{\rm vol}\,\text{SL}(2, \mathbb{C})}$, and the single-trace tree amplitude of $m$ gravitons and $r=n-m$ gluons is given by~\cite{Schlotterer:2016cxa}
\be\label{singletr}
\mathcal{M}_{m,r}^{\text{het}}(\rho)=\int | d\mu_n^{\rm string}|^2\,\text{PT}(\rho)R_m(z)\mathcal{K}_{n}(\bar z)\,,
\ee
where $\rho\in S_r/Z_r$ denotes the gluon color ordering. The holomorphic function $R_m$ contains half of the graviton vertex operator correlation function, taking the form
\begin{align}\label{R_m}
    R_m(z)=\sum_{(I)(J)\cdots(K)\in S_m}\mathcal{R}_{(I)}\mathcal{R}_{(J)}\cdots \mathcal{R}_{(K)}\,,
\end{align}
where we sum over permutations in $S_m$, each of which is written as a product of disjoint cycles. The cycle factor $\mathcal{R}_{(I)}$ is given by $\mathcal{R}_{(a)}=\sum_{b\neq a}\frac{\epsilon_a\Cdot k_b}{z_{ab}}:=C_a$, $\mathcal{R}_{(ab)}=\frac{\epsilon_a\Cdot\epsilon_b}{\alpha'z_{ab}^2}$
and $\mathcal{R}_{(I)}=0$ for $|I|>2$. The antiholomorphic function $\mathcal{K}_n (\bar{z})$ is the type-I correlator that encodes vertex operators of gluons and the other half of the gravitons. 

For the single-trace integrand, we can always gauge fix the gluon $z_n\rightarrow\infty$ such that we only need to break the $z$-cycles in $R_m(z)$ by IBP. The one-graviton integrand is automatically in the logarithmic form since $R_1=C_1$ contains no $z$-cycles, {\it i.e.} $\mathcal{I}_n(1;\rho)=\mathcal{I}'_n(1;\rho)=\text{PT}\,(\rho)\,C_1$\,.
For two gravitons, we have $R_2(z)=C_1C_2+\frac{\epsilon_1\Cdot\epsilon_2}{\alpha'z_{12}^2}$, which has only one $z$-cycle $\frac{1}{z_{12}^2}$. We can break it by the IBP
\begin{align}\label{2cycIBP}
    \frac{1}{z_{12}z_{21}}(\cdots)\overset{\rm IBP}{\cong}\sum_{i=3}^{n-1}\frac{s_{2i}}{1-s_{12}}\frac{1}{z_{12}z_{2i}}(\cdots)\,,
\end{align}
 which holds when $(\cdots)$ does not involve $z_2$.
This leads to the logarithmic function
\begin{align}\label{2hlog}
    \mathcal{I}'_n(12;\rho)&=\Bigg[\sum_{i=3}^{n-1}\frac{\epsilon_1\Cdot k_i\,z_{in}}{z_{1i}z_{1n}}\sum_{j=3}^{n-1}\frac{\epsilon_2\Cdot k_j\,z_{jn}}{z_{2j}z_{2n}}\\
    &\quad+\frac{\alpha'\epsilon_1\Cdot k_2\,\epsilon_2\Cdot k_1-\epsilon_1\Cdot\epsilon_2}{\alpha'(1-s_{12})z_{12}}\sum_{i=3}^{n-1}\frac{s_{2i}z_{in}}{z_{2i}z_{1n}}\Bigg]\text{PT}\,(\rho)\nonumber
\end{align}
with the SL$(2)$ covariance restored. As another example, we present the IBP of the three-graviton integrand in Appendix~\ref{sec:3hIBP}. Similar calculation can also be found in~\cite{Schlotterer:2016cxa} for a slightly different purpose. Although the complexity of $\mathcal{I}'_n$ grows quickly with more gravitons, it is still well controlled under our algorithm. Moreover, as the second step, we can greatly simplify $\mathcal{I}'_n$ by using SE and obtain a closed-form $\mathcal{I}''_n$ for the most generic single-trace sector.


Similar treatment can in principle be applied to multitrace and pure graviton sectors. We give several such examples in Appendix~\ref{sec:multitrace} and~\ref{sec:puregluon}.

\subsection{The single-trace integrand}\label{sec:stintegrand}
Now we present the single-trace CHY half-integrand $\mathcal{I}''_n$, resulted from our two-step reduction of string integrand~\eqref{singletr}:
\begin{align}\label{singletrCHY}
    \mathcal{I}''_n(12\cdots m;\rho)=\text{PT}\,(\rho)\mathcal{P}_m
\end{align}
where $\mathcal{P}_m$ is given by the cycle expansion
\begin{align}\label{P_m}
    \mathcal{P}_m=\sum_{(I)(J)\cdots(K)\in S_m}\Psi_{(I)}\Psi_{(J)}\cdots\Psi_{(K)}\,.
\end{align}
For length one and two cycles, we have $\Psi_{(a)}=C_a$ and $\Psi_{(ab)}=-T_{ab}\text{PT}(ab)$, where $T_{ab}$ is given by $T_{ab}:=\frac{\text{tr}\,(ab)}{1-s_{ab}}:=\frac{1}{2}\frac{\text{tr}\,(f_af_b)}{1-s_{ab}}$,
and $f_a^{\mu\nu}=k_a^\mu \epsilon_a^\nu-k_a^\nu\epsilon_a^\mu$ is the linearized field strength. For longer cycles ($|I|\geqslant 3$), we have $\Psi_{(I)}=-\frac{T_{I}\,\text{PT}\,(I)}{2}$, where $T_{I}$ is recursively generated from $T_{ab}$ through
\begin{align}\label{T_I}
    T_{I}=\frac{1}{1-s_I}\left[\text{tr}\,(I)+
    \sum_{\text{CP}} G_{i_2}^{I_1}G_{i_3}^{I_2}\cdots G_{i_1}^{I_p}\right]\,,
\end{align}
  with $\text{tr}(I):=\text{tr}(f_{a_1}f_{a_2}\cdots f_{a_{|I|}})$ for $|I|\geqslant 3$.
In this equation, the summation is over all cyclic partitions (CP) of a cycle $I$ into words, $\{I_1,I_2, \cdots, I_p\}$ where the length of each word $I_\ell$ is at least two; $i_\ell$ denotes the first element of $I_\ell$, which is the label in $I$ that succeeds the word $I_{\ell{-}1}$. For example, for $I=(1,2,3,4)$, we have four CPs with only one word, $1234$, $2341$, $3412$, $4123$, and two more with two words,  $\{12,34\}$ and $\{23,41\}$; the sum on the RHS of Eq.~\eqref{T_I} thus include terms like $G^{1234}_1$, and $G^{12}_3 G^{34}_1$, {\it etc}. Given a word $A=a_1a_2\cdots a_s$ and the next label $b$, the function $G$ is defined as
\begin{align}\label{G_I}
    G^{A}_b=\sum_{q=2}^{s-\delta_{a_1b}}T_{a_1a_2[a_3[\cdots[a_{q-1}\,a_q]\cdots]}V{}^{a_qb}_{a_{q+1}\cdots a_r}\,,
\end{align}
where  $V{}^{a_qb}_{a_{q+1}\cdots a_r}=\alpha'k_{a_q}\Cdot f_{a_{q+1}}\cdots f_{a_r}\Cdot k_b$, and in particular $V^{ab}=s_{ab}$. The bracket $[i\,j]$ stands for the index antisymmetrization, for example, $T_{ab[c\,d]}:= T_{abcd}-T_{abdc}$. The summation range $s-\delta_{a_1b}$ ensures that the right hand side of Eq.~\eqref{T_I} only contains those $T$'s with shorter cycles, since $b$ agrees with $a_1$ only when $A=I$. In Appendix~\ref{sec:examples}, we give examples of $T_I$ up to five gluons.

Note that the closed-formula for ${\cal I}_n''$ is still a conjecture, since it is generalized from explicit calculations at low points. However, as will be discussed in great detail in~\cite{Broedel:2013ttaa}, the formula has passed various nontrivial consistency checks. On the support of SE, we have checked numerically up to eleven points that Eq.~\eqref{singletrCHY} agrees with the logarithmic form $\mathcal{I}'_n$ resulted from the IBP of the string integrand~\eqref{singletr}. We have also checked up to eight points that the gauge amplitude produced by Eq.~\eqref{singletrCHY} agrees with that calculated from the Feynman diagrams of $(DF)^2+\text{YM}+\phi^3$ with massless external particles. Further consistency checks are provided in Appendix~\ref{sec:consistency}.

\section{Conclusion and discussion}

In this work we have presented an IBP reduction method for correlators of any massless string amplitude, ${\cal I}_n$, to a class of logarithmic functions, ${\cal I}_n'$. The latter can be used as CHY half-integrand (with the other half given by PT) to define field-theory amplitudes, and their double copy with disk/sphere integrals gives the original string amplitude. Although ${\cal I}_n'$ generally takes a complicated form, one can use SE to simplify it and even get a closed form to all multiplicities. We have demonstrated this two-step method for open (compactified) bosonic/heterotic strings, and obtained remarkably simple formulas for single-trace and certain multi-trace sectors of $(DF)^2+{\rm YM}+\phi^3$ theory. It can bring new insight into this theory, as well as conformal gravity which is obtained from a double copy with YM~\cite{Johansson:2017srf,Johansson:2018ues}.

Our discussion is completely generic since it only requires $\mathcal{I}_n$ to be a rational function with correct SL$(2)$ weight. It can thus be applied beyond those well-studied conventional string theories, for example, to the dual model proposed in Ref.~\cite{Baadsgaard:2016fel}. An important open question concerns the application of IBP reduction to string correlators with massive states, and it would be fascinating to see if similar structures exist there.

On the other hand, the CHY half-integrands obtained in this work contain explicit $\alpha'$ dependence. It will be interesting to understand how they can emerge from certain model of ambitwistor string~\cite{Mason:2013sva,Casali:2015vta,Geyer:2018xwu}. Some attempts have been made along this direction, and give the correct three-point results~\cite{Azevedo:2017yjy}. Last but not least, it is highly desirable to extend certain aspects of our construction to loop-level string amplitudes/correlators, and see how they are related to field-theory amplitudes/ambitwistor string correlators. 

\begin{acknowledgments}
It is our pleasure to thank Thales Azevedo, Henrik Johansson, Sebastian Mizera and Oliver Schlotterer for inspiring discussions and helpful comments on our draft.
S.H.'s research is supported in part by the Thousand Young Talents program, the Key Research Program of Frontier Sciences of CAS under Grant No. QYZDB-SSW-SYS014 and Peng Huanwu center under Grant No. 11747601.
F.T. is supported by the Knut and Alice Wallenberg Foundation under grant KAW 2013.0235, and the
Ragnar S\"{o}derberg Foundation under grant S1/16.
\end{acknowledgments}

\appendix

\section{More on IBP}\label{appa}  

In Sec.~\ref{sec:IBPproof}, we have proved that those gauge-fixed string correlators without numerators or connected subcycles can be reduced into logarithmic functions via IBP. Now we complete the proof for generic rational function $\mathcal{I}_n$ with correct SL$(2)$ weight by showing that we can always reduce them to the former case.
We start with an example  ${\cal I}_4=\frac{z_{13}z_{24}}{z_{12}^3z_{34}^3}$. We can absorb the numerator into the Koba-Nielsen factor as
\begin{align}\label{a3}
\frac{z_{13}z_{24}}{z_{12}^3z_{34}^3}\,\text{KN}=\frac{1}{z_{12}^2z_{34}^2}\frac{z_{13}z_{24}}{z_{12}z_{34}}\,\text{KN}:=\frac{1}{z_{12}^2z_{34}^2}\,\text{KN}'\,.
\end{align}
Now we can gauge fix $z_4\rightarrow\infty$ and perform the IBP as in Eq.~\eqref{I4IBP}, while the Mandelstam variables in $\text{KN}'$ are shifted as $u\rightarrow u+1$ and $s\rightarrow s-1$. We then restore the gauge covariance and release the cross ratio from $\text{KN}'$:
\begin{align}\label{a4}
\frac{1}{z_{12}^2z_{34}^2}\,\text{KN}'&\overset{\rm IBP}{\cong}\frac{t}{1-(s-1)}\text{PT}(1234)\,\text{KN}'\nonumber\\
&=\frac{t}{2-s}\left[\text{PT}(1234)\frac{z_{13}z_{24}}{z_{12}z_{34}}\right]\text{KN}\,.
\end{align}
The product in the bracket can be rearranged into a sum of rational functions without $z_{ij}$ in the numerators:
\begin{align}\label{a5}
{\rm PT}(1234) \frac{z_{13}z_{24}}{z_{12}z_{34}}&=
{\rm PT}(1234) \frac{(z_{12}+z_{23})z_{24}}{z_{12}z_{34}}
\nl
&={\rm PT}(1234)-{\rm PT}(12) {\rm PT}(34)\,,
\end{align}
Although the second term is not a logarithmic function, we have reduced the problem to the one already solved in Eq.~\eqref{I4IBP}: by using this IBP relation again, we obtain the final logarithmic function:
\begin{align}\label{a6}
\frac{z_{13}z_{24}}{z_{12}^3z_{34}^3}\overset{\rm IBP}{\cong}\frac{(1-s-t)\,t}{(1-s)(2-s)}\,\text{PT}(1234)\,,
\end{align}
which involves exotic tachyon poles. 


Our example captures the essence of the general idea. In any SL$(2)$ weight two integrand $\mathcal{I}_n$, if $z_{ab}$ appear in the numerator, $a$ and $b$ must also appear in the denominator in, say, $z_{ac}$ and $z_{bd}$.
We can remove the numerator $z_{ab}$ by absorbing the cross ratio $\frac{z_{ab}z_{cd}}{z_{ac}z_{bd}}$ into the Koba-Nielsen factor, schematically shown as
\begin{align*}
\begin{tikzpicture}[baseline={([yshift=-0.7ex]current bounding box.center)},every node/.style={font=\footnotesize},scale=1.25]
\draw (-0.5,0.25) -- (0.5,0.25) (-0.5,-0.25) -- (0.5,-0.25);
\draw [densely dashed] (-0.25,0.25) -- (-0.25,-0.25);
\filldraw (-0.25,0.25) circle (1pt) (-0.25,-0.25) circle (1pt) (0.25,0.25) circle (1pt) (0.25,-0.25) circle (1pt);
\draw (-0.25,0.25) -- ++(60:0.25) (-0.25,-0.25) -- ++(-60:0.25) (0.25,0.25) -- ++(120:0.25) (0.25,-0.25) -- ++(-120:0.25);
\foreach \x in {90,120,150} {
	\filldraw (-0.25,0.25) ++(\x:0.25) circle (0.3pt) (-0.25,-0.25) ++(-\x:0.25) circle (0.25pt);
}
\foreach \x in {30,60,90} {
	\filldraw (0.25,0.25) ++(\x:0.25) circle (0.3pt) (0.25,-0.25) ++(-\x:0.25) circle (0.25pt);
}
\node at (-0.25,0.25) [below left=-1.5pt]{$a$};
\node at (-0.25,-0.25) [above left=-1.5pt]{$b$};
\node at (0.25,0.25) [below right=-1.5pt]{$c$};
\node at (0.25,-0.25) [above right=-1.5pt]{$d$};
\path [decoration={markings, mark=at position \halfway with {\arrow{>}}},postaction={decorate}] (-0.25,0.25) -- (-0.25,-0.25);
\path [decoration={markings, mark=at position \halfway with {\arrow{>}}},postaction={decorate}] (-0.25,-0.25) -- (0.25,-0.25);
\path [decoration={markings, mark=at position \halfway with {\arrow{>}}},postaction={decorate}] (-0.25,0.25) -- (0.25,0.25);
\end{tikzpicture}\,\text{KN}=\begin{tikzpicture}[baseline={([yshift=-0.7ex]current bounding box.center)},every node/.style={font=\footnotesize},scale=1.25]
\draw (-0.5,0.25) -- (-0.25,0.25) (-0.5,-0.25) -- (-0.25,-0.25) (0.25,0.25) -- (0.5,0.25) (0.25,-0.25) -- (0.5,-0.25);
\draw (0.25,0.25) -- (0.25,-0.25);
\filldraw (-0.25,0.25) circle (1pt) (-0.25,-0.25) circle (1pt) (0.25,0.25) circle (1pt) (0.25,-0.25) circle (1pt);
\draw (-0.25,0.25) -- ++(60:0.25) (-0.25,-0.25) -- ++(-60:0.25) (0.25,0.25) -- ++(120:0.25) (0.25,-0.25) -- ++(-120:0.25);
\foreach \x in {90,120,150} {
	\filldraw (-0.25,0.25) ++(\x:0.25) circle (0.3pt) (-0.25,-0.25) ++(-\x:0.25) circle (0.25pt);
}
\foreach \x in {30,60,90} {
	\filldraw (0.25,0.25) ++(\x:0.25) circle (0.3pt) (0.25,-0.25) ++(-\x:0.25) circle (0.25pt);
}
\path [decoration={markings, mark=at position \halfway with {\arrow{>}}},postaction={decorate}] (0.25,0.25) -- (0.25,-0.25);
\node at (-0.25,0.25) [below left=-1.5pt]{$a$};
\node at (-0.25,-0.25) [above left=-1.5pt]{$b$};
\node at (0.25,0.25) [below right=-1.5pt]{$c$};
\node at (0.25,-0.25) [above right=-1.5pt]{$d$};
\begin{scope}[xshift=0.7cm]
\draw (0,0.25) node [below right=-1.5pt]{$a$} -- (0.5,0.25) node[below left=-1.5pt]{$c$} (0,-0.25) node[above right=-1.5pt]{$b$} -- (0.5,-0.25) node[above left=-1.5pt]{$d$};
\filldraw (0,0.25) circle (1pt) (0.5,0.25) circle (1pt) (0,-0.25) circle (1pt) (0.5,-0.25) circle (1pt);
\draw [densely dashed] (0,0.25) -- (0,-0.25) (0.5,0.25) -- (0.5,-0.25);
\path [decoration={markings, mark=at position \halfway with {\arrow{>}}},postaction={decorate}] (0.5,0.25) -- (0.5,-0.25);
\path [decoration={markings, mark=at position \halfway with {\arrow{>}}},postaction={decorate}] (0,0.25) -- (0,-0.25);
\path [decoration={markings, mark=at position \halfway with {\arrow{>}}},postaction={decorate}] (0,-0.25) -- (0.5,-0.25);
\path [decoration={markings, mark=at position \halfway with {\arrow{>}}},postaction={decorate}] (0,0.25) -- (0.5,0.25);
\end{scope}
\end{tikzpicture}\,\text{KN}=\begin{tikzpicture}[baseline={([yshift=-0.7ex]current bounding box.center)},every node/.style={font=\footnotesize},scale=1.25]
\draw (-0.5,0.25) -- (-0.25,0.25) (-0.5,-0.25) -- (-0.25,-0.25) (0.25,0.25) -- (0.5,0.25) (0.25,-0.25) -- (0.5,-0.25);
\draw (0.25,0.25) -- (0.25,-0.25);
\filldraw (-0.25,0.25) circle (1pt) (-0.25,-0.25) circle (1pt) (0.25,0.25) circle (1pt) (0.25,-0.25) circle (1pt);
\draw (-0.25,0.25) -- ++(60:0.25) (-0.25,-0.25) -- ++(-60:0.25) (0.25,0.25) -- ++(120:0.25) (0.25,-0.25) -- ++(-120:0.25);
\foreach \x in {90,120,150} {
	\filldraw (-0.25,0.25) ++(\x:0.25) circle (0.3pt) (-0.25,-0.25) ++(-\x:0.25) circle (0.25pt);
}
\foreach \x in {30,60,90} {
	\filldraw (0.25,0.25) ++(\x:0.25) circle (0.3pt) (0.25,-0.25) ++(-\x:0.25) circle (0.25pt);
}
\path [decoration={markings, mark=at position \halfway with {\arrow{>}}},postaction={decorate}] (0.25,0.25) -- (0.25,-0.25);
\node at (-0.25,0.25) [below left=-1.5pt]{$a$};
\node at (-0.25,-0.25) [above left=-1.5pt]{$b$};
\node at (0.25,0.25) [below right=-1.5pt]{$c$};
\node at (0.25,-0.25) [above right=-1.5pt]{$d$};
\end{tikzpicture}\,\text{KN}'\,.
\end{align*}
By repeating this operation, we obtain an integrand that contains only disjoint subcycles. After gauge fixing, we can then follow Sec.~\ref{sec:IBPproof} and reduce it to a sum of PT factors via IBP relations, like the first row of Eq.~\eqref{a4}.



Next, we release a cross ratio, say, $\frac{z_{ab}z_{cd}}{z_{ac}z_{bd}}$ from the deformed Koba-Nielsen factor $\text{KN}'$. For each PT factor, we first use KK relations to move $b$ next to $a$, and then gauge fix $z_c\rightarrow\infty$. The resultant integrand only contains a tadpole and a labeled line:  
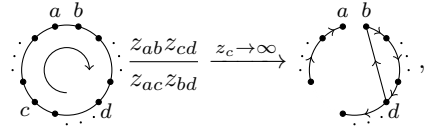
\begin{align*}
\begin{tikzpicture}[baseline={([yshift=-0.7ex]current bounding box.center)},every node/.style={font=\footnotesize}]
\draw (0,0) circle (0.6cm);
\filldraw (45:0.6) circle (1pt) (75:0.6) circle (1pt) node[above=0pt]{$b$} (105:0.6) circle (1pt) node[above=0pt]{$a$} (135:0.6) circle (1pt) (-45:0.6) circle (1pt) node[below right=-3pt]{$d$} (-135:0.6) circle (1pt) node[below left=-1.5pt]{$c$} (-15:0.6) circle (1pt) (-165:0.6) circle (1pt) (-105:0.6) circle (1pt);
\foreach \x in {180,165,150,30,15,0,-60,-75,-90} {
	\filldraw (\x:0.72) circle (0.2pt);
}
\draw [->] (270:0.3) arc (270:0:0.3);
\end{tikzpicture}\,\frac{z_{ab}z_{cd}}{z_{ac}z_{bd}}
\xrightarrow{z_c\rightarrow\infty}\,\begin{tikzpicture}[baseline={([yshift=-0.7ex]current bounding box.center)},every node/.style={font=\footnotesize}]
\draw (0,0) circle (0.6cm);
\draw [white,very thick] (75:0.6) arc (75:105:0.6) (-105:0.6) arc (-105:-165:0.6);
\draw [decoration={markings, mark=at position \halfway with {\arrow{>}}},postaction={decorate}] (-45:0.6) -- (75:0.6);
\filldraw (-105:0.6) circle (1pt);
\filldraw (45:0.6) circle (1pt) (75:0.6) circle (1pt) node[above=0pt]{$b$} (105:0.6) circle (1pt) node[above=0pt]{$a$} (135:0.6) circle (1pt) (-45:0.6) circle (1pt)  node[below right=-3pt]{$d$}  (-15:0.6) circle (1pt) (-165:0.6) circle (1pt);
\foreach \x in {180,165,150,30,15,0,-60,-75,-90} {
	\filldraw (\x:0.72) circle (0.2pt);
}
\path [decoration={markings, mark=at position \halfway with {\arrow{>}}},postaction={decorate}] (75:0.6) arc (75:45:0.6);
\path [decoration={markings, mark=at position 0.9*\halfway with {\arrow{>}}},postaction={decorate}] (45:0.6) arc (45:-15:0.6);
\path [decoration={markings, mark=at position \halfway with {\arrow{>}}},postaction={decorate}] (-15:0.6) arc (-15:-45:0.6);
\path [decoration={markings, mark=at position 0.9*\halfway with {\arrow{>}}},postaction={decorate}] (-45:0.6) arc (-45:-105:0.6);
\path [decoration={markings, mark=at position 0.9*\halfway with {\arrow{>}}},postaction={decorate}] (195:0.6) arc (195:135:0.6);
\path [decoration={markings, mark=at position 0.9*\halfway with {\arrow{>}}},postaction={decorate}] (135:0.6) arc (135:105:0.6);
\end{tikzpicture}~,
\end{align*}
which can be turned into a sum of PT factors via IBP according to Sec.~\ref{sec:IBPproof}. This process is captured by Eq.~\eqref{a5} and~\eqref{a6} in our example. By recursively releasing cross ratios from the deformed Koba-Nielsen factor, we will eventually land on a logarithmic function together with the original string measure, which completes our constructive proof.

\section{Single trace integrand with three gravitons}\label{sec:3hIBP}
In this section, we derive the three-graviton CHY integrand from IBP. This calculation is similar to the one in Ref.~\cite{Schlotterer:2016cxa}, but we present the result in a different form. We start with $R_3(z)$, given by Eq.~\eqref{R_m}:
\begin{align}\label{R_3}
R_3(z)&=C_1C_2C_3\nonumber\\
&\quad+\frac{\epsilon_1\Cdot\epsilon_2}{\alpha'z_{12}^2}C_3+\frac{\epsilon_1\Cdot\epsilon_3}{\alpha'z_{13}^2}C_2+\frac{\epsilon_2\Cdot\epsilon_3}{\alpha'z_{23}^2}C_1\,,
\end{align}
and we work with the gauge fix $z_n\rightarrow\infty$. 

Our first step is to separate out the length-two and three $z$-cycles contained in $C_1C_2C_3$ and combine them with the second row of Eq.~\eqref{R_3}:
\begin{align}\label{R_3step1}
R_3(z)&=C_1C_2C_3-C_{12}C_{21}C_3-C_{13}C_{31}C_2-C_{23}C_{32}C_1\nonumber\\
&\quad-C_{12}C_{23}C_{31}-C_{13}C_{32}C_{21}\nonumber\\[-7pt]
\cdashline{2-2}
&\quad+\frac{(1-s_{12})\mathcal{E}_{12}C_3}{\alpha'z_{12}^2}+\frac{(1-s_{13})\mathcal{E}_{13}C_2}{\alpha'z_{13}^2}\nonumber\\
&\quad+\frac{(1-s_{23})\mathcal{E}_{23}C_1}{\alpha'z_{23}^2}+C_{12}C_{23}C_{31}+C_{13}C_{32}C_{21}\,,
\end{align}
where we have defined $C_{ij}:=\frac{\epsilon_i\Cdot k_j}{z_{ij}}$ and
\begin{align}
\mathcal{E}_{ij}=\frac{\epsilon_i\Cdot\epsilon_j-\alpha'\epsilon_i\Cdot k_j\,\epsilon_j\Cdot k_i}{1-s_{ij}}=\epsilon_i\Cdot\epsilon_j-\alpha'T_{ij}\,.
\end{align}
In Eq.~\eqref{R_3step1}, the two rows above the dashed line do not contain any $z$-cycles, as one can check straightforwardly. The first three terms below the dashed line of Eq.~\eqref{R_3step1} contain only length-two $z$-cycles, while the last two terms contain only length-three cycles. All the length-two cycles can be broken by Eq.~\eqref{2cycIBP} and the following relation:
\begin{align}
\frac{1}{z_{12}z_{21}}\frac{1}{z_{31}}\overset{\rm IBP}{\cong} \frac{1}{1-s_{12}}\Bigg[\sum_{i=4}^{n-1}\frac{s_{2i}}{z_{2i}z_{31}z_{12}}+\frac{s_{23}}{z_{12}z_{23}z_{31}}\Bigg]\,,
\end{align}
which holds when its coefficient does not contain $z_2$. We note that this IBP generates new length-three cycles. 

Next, we can break all the length-three cycles by the IBP relation
\begin{align}
\frac{1}{z_{12}z_{23}z_{31}}\overset{\rm IBP}{\cong}\frac{1}{1-s_{123}}\sum_{i=4}^{n-1}\Bigg[\frac{s_{3i}}{z_{12}z_{23}z_{3i}}-\frac{s_{2i}}{z_{13}z_{32}z_{2i}}\Bigg]\,,
\end{align}
which holds when its coefficient does not contain either $z_2$ or $z_3$. The final result is
\begin{widetext}
	\begin{align}\label{3hlogform}
	R_3(z)&\overset{\rm IBP}{\cong} C_1C_2C_3-C_{12}C_{21}C_3-C_{13}C_{31}C_2-C_{23}C_{32}C_1-C_{12}C_{23}C_{31}-C_{13}C_{32}C_{21}\nonumber\\
	&\quad-\Bigg[\frac{\mathcal{E}_{12}}{z_{12}}\sum_{i=3}^{n-1}\frac{k_2\Cdot k_i}{z_{2i}}\sum_{j=4}^{n-1}\frac{\epsilon_3\Cdot k_j}{z_{3j}}+\frac{\epsilon_3\Cdot k_1\,\mathcal{E}_{12}}{z_{31}z_{12}}\sum_{i=4}^{n-1}\frac{k_2\Cdot k_i}{z_{2i}}+\sum_{i=4}^{n-1}\frac{k_1\Cdot k_i}{z_{1i}}\frac{\epsilon_3\Cdot k_2\,\mathcal{E}_{21}}{z_{32}z_{21}}+\text{cyc}(123)\Bigg]\nonumber\\
	&\quad+\frac{T_{123}}{2}\sum_{i=4}^{n-1}\Bigg[\frac{s_{3i}}{z_{12}z_{23}z_{3i}}-\frac{s_{2i}}{z_{13}z_{32}z_{2i}}\Bigg]+(2\leftrightarrow 3)\,.
	\end{align}
\end{widetext}
After being multiplied by a PT factor, Eq.~\eqref{3hlogform} is just the logarithmic form integrand $\mathcal{I}'_n$ for the three-graviton case. 

Since $\mathcal{I}'_n$ is also a CHY integrand, we can use SE to rewrite it into a more compact form. For example, the summation in the last row of Eq.~\eqref{3hlogform} can be reduced to
\begin{align}
\sum_{i=4}^{n-1}\Bigg[\frac{s_{3i}}{z_{12}z_{23}z_{3i}}-\frac{s_{2i}}{z_{13}z_{32}z_{2i}}\Bigg]=-\frac{s_{123}}{z_{12}z_{23}z_{31}}\,.
\end{align}
One can easily show that on the support of SE, Eq.~\eqref{3hlogform} agrees with the $\mathcal{P}_3$ given in the next section.
For generic cases, the operation resembles the inverse one as shown in~\cite{Cardona:2016gon,Bjerrum-Bohr:2016axv,Huang:2017ydz}.

\section{Single-trace examples}\label{sec:examples}
Now we provide a few examples for our generic single-trace formula given in Sec.~\ref{sec:stintegrand}. The integrand involving two gravitons is given by $\text{PT}(\rho)\mathcal{P}_2$, where 
\begin{align}
\mathcal{P}_2=C_1C_2+\Psi_{(12)}=C_1C_2-\frac{\text{tr}\,(12)\,\text{PT}\,(12)}{1-s_{12}}\,,
\end{align}
which agrees with the logarithmic function~\eqref{2hlog} after some use of SE. For three gravitons, Eq.~\eqref{P_m} gives
\begin{align}\label{P_3}
\mathcal{P}_3&=C_1C_2C_3+\left(C_1\Psi_{(23)}+C_2\Psi_{(13)}+C_3\Psi_{(12)}\right)\nonumber\\
&\quad+\Psi_{(123)}+\Psi_{(132)}\nonumber\\
&=C_1C_2C_3-\left[\frac{C_1\text{tr}\,(23)\,\text{PT}\,(23)}{1-s_{23}}+\text{cyc}(123)\right]\nonumber\\
&\quad-\frac{1}{2}\left[T_{123}\text{PT}\,(123)+T_{132}\text{PT}\,(132)\right]\,,
\end{align}
where, according to Eq.~\eqref{T_I}, $T_{123}$ is given by
\begin{align}\label{T_123}
T_{123}&=\frac{1}{1-s_{123}}\left[\text{tr}\,(123)+G^{123}_1+G^{231}_2+G^{312}_3\right]\nl
&=\frac{\text{tr}\,(123)}{1-s_{123}}+{\alpha'}\!\left[\frac{\text{tr}(12)(k_2\Cdot f_3\Cdot k_1)}{(1-s_{12})(1-s_{123})}+\text{cyc}(123)\right]\nonumber
\end{align}
In the four-graviton integrand $\mathcal{P}_4$, the only piece that has not appeared in $\mathcal{P}_2$ or $\mathcal{P}_3$ is $T_{abcd}$ in the length-four cycle factor $\Psi_{(abcd)}$, which is given by
\begin{align}
T_{1234}&=
\frac{\text{tr}\,(1234)+\!\left[T_{12}V{}^{21}_{34} +T_{123}V{}^{31}_4+\text{cyc}(1234)\right]}{1-s_{1234}}\nonumber\\
&\quad+\frac{T_{12}s_{23}T_{34}s_{41}+T_{23}s_{34}T_{41}s_{12}}{1-s_{1234}}\,.
\end{align}
The first example of index antisymmetrization for $G$ function as shown in Eq.~\eqref{G_I} is
\begin{align*}
G^{12345}_{1}&= T_{12}V{}^{21}_{345} + T_{123}V{}^{31}_{45}+ (T_{1234}-T_{1243}) V{}^{41}_5\,.
\end{align*}
It is contained in
\begin{align}
T_{12345}&=\frac{\text{tr}\,(12345)}{1-s_{12345}}+\frac{G^{12345}_1+\text{cyc}(12345)}{1-s_{12345}}\\
&\quad+\frac{T_{12}s_{23}(T_{34}V{}^{41}_5+T_{345}s_{51})+\text{cyc}(12345)}{1-s_{12345}}\nonumber
\end{align}
that appears in the length-five cycle factor $\Psi_{(12345)}$.

\section{Consistency checks}\label{sec:consistency}

Last but not least, we list some all-multiplicity checks. From Eq.~\eqref{T_I} and~\eqref{G_I}, one can see that $T_I$ satisfies the same cyclic and reflection property as $\text{tr}\,(I)$. Gauge invariance is manifest since $T_I$ is built from $f$'s and $C_a$ is gauge invariant on the support of SE. Moreover, in the limit $k_m\rightarrow 0$, we have $T_I\rightarrow 0$ if $m\in I$, such that $\mathcal{P}_m$ factorizes as $\mathcal{P}_m\rightarrow C_m\mathcal{P}_{m-1}$, recovering the universal single soft behavior~\cite{Weinberg:1965nx} after the CHY integration.  Furthermore, as $\alpha'\rightarrow 0$, all the tachyon poles $(1-s_I)^{-1}$ drop out, and by induction it is easy to see that all the $G$-factors vanish as $\mathcal{O}(\alpha')$. Consequently, $\mathcal{P}_m$ reduces to the cycle expansion of a Pfaffian~\cite{Lam:2016tlk}, reproducing the single-trace YM-scalar integrand $\text{PT}(\rho)\,\text{Pf}\,\pmb{\Psi}_m$~\cite{Cachazo:2014nsa}.
On the other hand, the $T_I$'s vanish as $\mathcal{O}(\alpha'^{-1})$ when $\alpha'\rightarrow\infty$, leaving only $C_a$'s in $\mathcal{P}_m$, which lands exactly on the $(DF)^2+\phi^3$ integrand $\text{PT}\,(\rho)(\prod_{i=1}^m C_i)$~\cite{Azevedo:2017lkz}. 

The most important check is to show that the integrand given by~\eqref{singletrCHY} has the correct massless factorization properties for the $(DF)^2+\text{YM}+\phi^3$ amplitudes~\cite{Broedel:2013ttaa}. Following~\cite{Cachazo:2013iea}, $d\mu^{\text{CHY}}_n$ and PT's split correspondingly in the factorization limit $p^2\rightarrow 0$, thus we only need to consider the behavior of ${\cal P}_m$. The scalar channel, which separates the gluons into two sets, is simple: the leading order gives a simple pole $\frac{1}{p^2}$, and $\mathcal{P}_m$ factorizes nicely since all $\Psi_{(I)}$'s that contain gluons in both sets only contribute to higher orders. In other words, the cycle expansion form shown in Eq.~\eqref{P_m} trivializes the scalar factorization channel. 

The gluon channel is more interesting. For simplicity, we consider the special one that separates the gluons and the scalars. The leading order is now a double pole $\frac{1}{p^4}$ whose coefficient is proportional to $\mathcal{P}_m$ evaluated on the $(m+1)$-point kinematics, while the subleading order provides the simple pole $\frac{1}{p^2}$. The leading order must vanish, and indeed we have checked up to eight points that
\begin{align}\label{corank2}
\mathcal{P}_m=0\quad\text{for $m$ and $(m+1)$-point kinematics.}
\end{align}
When $\alpha'=0$, this statement is simply that the matrix $\pmb{\Psi}_m$ has corank two for $m$-point kinematics. Remarkably, $\mathcal{P}_m$ inherits this corank-two property for finite $\alpha'$! One can verify this statement by showing that the coefficient of each tachyon pole structure of $\mathcal{P}_m$ vanishes on the support of the SEs, as we will present in~\cite{Broedel:2013ttaa}.

\section{Towards multi-trace integrands}\label{sec:multitrace}
The generalization of the single-trace string integrand
to $t$ color traces is straightforward:
\begin{align}\label{multitr}
\mathcal{M}_{m,r}^{\text{het}}(\{\rho_i\})=\!\int\! | d\mu_n^{\rm string}|^2\Bigg[\prod_{i=1}^{t}\text{PT}(\rho_i)\Bigg]R_m(z)\mathcal{K}_{n}(\bar z)\,,
\end{align}
where $\rho_i\in S_{r_i}/Z_{r_i}$ and $\sum_{i=1}^t r_i=r$. Following the same prescription outlined in Sec.~\ref{sec:heterotic}, we can in principle get the CHY integrand $\mathcal{I}''_n$ for multitrace sectors.

Interestingly, we can also get multitrace $\mathcal{I}''_n$ by taking certain coefficients of the single-trace integrand $\mathcal{I}''_n(h;\rho)$. As the most general formula will be given in~\cite{Broedel:2013ttaa}, here we simply present it for some cases. To start with, the double-trace integrand can be obtained by the prescription
\begin{align}\label{takecoe}
\renewcommand{\arraystretch}{1.1}
\mathcal{I}''_n(h';\sigma,\rho)=\left(\begin{array}{c}
\text{the coefficient of} \\ 
\epsilon_{\sigma_1}\Cdot k_{\sigma_2}\epsilon_{\sigma_2}\Cdot k_{\sigma_3}\cdots\epsilon_{\sigma_{|\sigma|}}\Cdot k_{\sigma_1} \\
\text{ in }\mathcal{I}''_n(h;\rho)
\end{array}\right)\,,
\end{align}
where $\sigma\subseteq h$ denotes another gluon trace (in addition to $\rho$) and $h'=h\backslash\sigma$ denotes the new set of gravitons. The simplest one is the pure-gluon double trace case already given in~\cite{Schlotterer:2016cxa,Azevedo:2018dgo}: ${\cal I}''_n( \sigma,\rho)=-{\rm PT}(\sigma) {\rm PT}(\rho) \frac{s_{\sigma}}{1-s_{\sigma}}$. We also give a closed-form expression for double trace with one additional graviton, denoted by label $q$:
\begin{align}\label{dtr}
\mathcal{I}''_n(q;\sigma,\rho)=&\frac{\text{PT}(\rho)}{1-s_\sigma}\Bigg[-s_\sigma\text{PT}(\sigma)C_q-\frac{\alpha'}{2(1-s_\rho)}\\
&\times\!\!\!\!\!\!\!\sum_{\substack{\sigma_i,\sigma_j\in\sigma \\ \gamma\in \{\sigma_i\,X\shuffle Y^T\,\sigma_j\}}}\!\!\!\!\!\!\!(-1)^{|Y|}  k_{\sigma_j}\Cdot f_q\cdot \!k_{\sigma_i}\text{PT}(\gamma,q)\Bigg],\nonumber
\end{align}
where we first sum over all pairs of labels $\sigma_i, \sigma_j$, and for each choice we read out words $X$ and $Y$ by rewriting the cycle $\sigma$ in the form $(\sigma_i,X,\sigma_j,Y)$; then the sum is over words $\gamma$ with endpoints $\gamma_1=\sigma_i$ and $\gamma_{|\gamma|}=\sigma_j$, with shuffle of $X$ and $Y^T$ in between. On the support of SEs, Eq.~\eqref{dtr} is invariant under the relabeling of the traces, although this property is not explicit%
.

Furthermore, we can extract a third gluon trace $\tau$ from the double-trace integrand $\mathcal{I}''_n(h';\sigma,\rho)$ by the same process of Eq.~\eqref{takecoe}. Remarkably, the pure-gluon triple-trace result also takes a compact form:
\begin{align}\label{ttr}
\mathcal{I}''_n(\tau,\sigma,\rho)&=\frac{\text{PT}(\rho)}{(1-s_\tau)(1-s_\sigma)}\Big[s_\tau s_\sigma\text{PT}(\tau)\text{PT}(\sigma)\nonumber\\
&\quad-\frac{1}{2(1-s_\rho)}\!\!\sum_{\substack{\tau_k,\tau_\ell\in\tau \\ \beta \in\{\tau_k\,Z\shuffle W^T\tau_\ell\}}}\sum_{\substack{\sigma_i,\sigma_j\in\sigma \\ \gamma\in\{\sigma_i\,X\shuffle Y^T\sigma_j\}}}\nonumber\\
&\quad\times(-1)^{|W|+|Y|}s_{\tau_\ell \sigma_i}s_{\sigma_j\tau_k}\text{PT}(\beta,\gamma)\Big]\,.
\end{align}

At the leading order of $\alpha'$, both Eq.~\eqref{dtr} and~\eqref{ttr} reproduce the YM-scalar theory in the corresponding multitrace sectors~\cite{Cachazo:2014xea,Chiodaroli:2014xia,Chiodaroli:2017ngp}.

We have checked that Eq.~\eqref{dtr} and~\eqref{ttr} agree with IBP result starting from Eq.~\eqref{multitr} on the support of SE. We can get generic multitrace integrands by repeating Eq.~\eqref{takecoe}, which has been numerically checked up to eleven points and five traces with IBP results.

We provide two examples for Eq.~\eqref{dtr} and~\eqref{ttr}, both of which are at six points:
\begin{align}
\mathcal{I}''_6(1;23,456)
&=-\frac{s_{23}}{1-s_{23}}C_1\text{PT}(23)\text{PT}(456)\nonumber\\
&\quad-\frac{\alpha'(k_3\Cdot f_1\Cdot k_2)\text{PT}(123)\text{PT}(456)}{(1-s_{23})(1-s_{456})}\,,
\end{align}
\begin{align}
\mathcal{I}''_6(12,34,56)
&=\text{PT}(56)\left[\frac{s_{12}s_{34}\text{PT}(12)\text{PT}(34)}{(1-s_{12})(1-s_{34})}\right.\\
&\quad-\left.\frac{s_{23}s_{41}\text{PT}(1234)+s_{24}s_{31}\text{PT}(1243)}{(1-s_{12})(1-s_{34})(1-s_{56})}\right].\nonumber
\end{align}

\section{Pure gluon integrand}\label{sec:puregluon}
It turns out to be convenient to write pure $(DF)^2+\text{YM}$ integrand $\mathcal{I}''_n=\text{PT}(12 \cdots n)\mathcal{P}'_n$ as
\begin{align}
\mathcal{P}'_n=(-1)^{\frac{n(n-1)}{2}}\text{Pf}\,'(\pmb{\Psi}_n)+\alpha'\mathcal{Q}_n\,,
\end{align}
where the second term vanishes at $\alpha'=0$. Moreover, by taking $\alpha'=0$ in $\mathcal{Q}_n$, we should reproduce the half-integrand for a single insertion of $F^3$ vertex~\cite{He:2016iqi}. This instructs us to use a similar cycle expansion for $\mathcal{Q}_n$: 
\begin{align}\label{T'_n}
\mathcal{Q}_n=\sum_{(I)(J)\cdots(K)\in S_n}(N_{>1}-1)
\langle{\Psi_{(I)}\Psi_{(J)}\cdots\Psi_{(K)}}\rangle\,,
\end{align}
where $N_{>1}$ counts the number of cycles with length at least two. The bracket $\langle\cdots\rangle$ stands for certain ``averaging'' over cycles with length at least two, which will be spelled out shortly, while length-one cycles factor out as:
\begin{align}
\langle{\Psi_{(a)}\Psi_{(J)}\cdots\Psi_{(K)}}\rangle=C_a\langle{\Psi_{(J)}\cdots\Psi_{(K)}}\rangle\,.
\end{align}
The longest cycles involved in $\mathcal{Q}_n$ have length $n-2$.

At four points, the $(DF)^2+\text{YM}$ amplitude is explicitly given in~\cite{Azevedo:2018dgo}, which can be produced by
\begin{align}
\mathcal{Q}_4=-C_1C_2C_3C_4 &+\langle{\Psi_{(12)}\Psi_{(34)}}\rangle+\langle{\Psi_{(23)}\Psi_{(41)}}\rangle\nonumber\\
&+\langle\Psi_{(13)}\Psi_{(24)}\rangle\,,
\end{align}
where the last three terms are given by the relabeling of
\begin{align}\label{Psi12Psi34}
\langle{\Psi_{(12)}\Psi_{(34)}}\rangle=\frac{1}{2}\frac{T_{12}\text{tr}(34)+\text{tr}(12)T_{34}}{z_{12}^2z_{34}^2}\,.
\end{align}
When going to higher points, the single soft limit fixes all the terms involving length-one cycles, while the requirement of no double pole at massless factorization channels puts strong constraints on the terms without length-one cycles. Actually, we find that up to seven points these constraints lead to a unique answer to the ansatz~\eqref{T'_n}, which involves a modified version of $T_I$:
\begin{align}
\tilde{T}_{I}=\frac{1}{1-s_I}\left[\text{tr}\,(I)+\sum_{\text{CP}}g_p G_{i_2}^{I_1}G_{i_3}^{I_2}\cdots G_{i_1}^{I_p}\right]\,,
\end{align}
where $g_p$ depends on $p$, the number of partitions. For $n\leqslant 7$, we always have $|I|\leqslant 5$, such that the only relevant $g_p$'s are $g_1=1$ and $g_2=\frac{2}{3}$. In the following, we give explicitly the building blocks of $\mathcal{Q}_n$ up to seven points.

\paragraph{Five points:} Besides Eq.~\eqref{Psi12Psi34}, we also need
\begin{align}\label{Psi12Psi345}
\langle{\Psi_{(12)}\Psi_{(345)}}\rangle=\frac{1}{4}\frac{T_{12}\text{tr}(345)+\text{tr}(12)T_{345}}{z_{12}z_{21}z_{34}z_{45}z_{53}}\,.
\end{align}
The overall normalization here is different from that in Eq.~\eqref{Psi12Psi34} because of the different normalizations adopted for $T_{ab}$ and longer cycles.

\paragraph{Six points: } In addition to Eq.~\eqref{Psi12Psi34} and~\eqref{Psi12Psi345}, there are three more building blocks:
\begingroup
\allowdisplaybreaks
\begin{align}
&\langle{\Psi_{(12)}\Psi_{(3456)}}\rangle=\frac{1}{4}\frac{T_{12}\text{tr}(3456)+\text{tr}(12)\tilde{T}_{3456}}{z_{12}z_{21}z_{34}z_{45}z_{56}z_{63}}\,,\\
&\langle{\Psi_{(123)}\Psi_{(456)}}\rangle=\frac{1}{8}\frac{T_{123}\text{tr}(456)+\text{tr}(123)T_{456}}{z_{12}z_{23}z_{31}z_{45}z_{56}z_{64}}\,,\\
&\langle{\Psi_{(12)}\Psi_{(34)}\Psi_{(56)}}\rangle=\frac{1}{6}\frac{1}{z_{12}^2z_{34}^2z_{56}^2}\nonumber\\*
&\qquad\qquad\times\Big[T_{12}\text{tr}(34)\text{tr}(56)+T_{34}\text{tr}(56)\text{tr}(12)\nonumber\\*
&\qquad\qquad\quad\;+T_{56}\text{tr}(12)\text{tr}(34)+T_{12}T_{34}\text{tr}(56)\nonumber\\*
&\qquad\qquad\quad\;+T_{34}T_{56}\text{tr}(12)+T_{56}T_{12}\text{tr}(34)\Big]\,.
\end{align}
\endgroup

\paragraph{Seven points: } There are again three new building blocks emerge at $n=7$:
\begingroup
\allowdisplaybreaks
\begin{align}
&\langle{\Psi_{(123)}\Psi_{(4567)}}\rangle=\frac{1}{8}\frac{T_{123}\text{tr}(4567)+\text{tr}(123)\tilde{T}_{4567}}{z_{12}z_{23}z_{31}z_{45}z_{56}z_{67}z_{74}}\,,\\
&\langle{\Psi_{(12)}\Psi_{(34567)}}\rangle=\frac{1}{4}\frac{T_{12}\text{tr}(34567)+\text{tr}(12)\tilde{T}_{34567}}{z_{12}z_{21}z_{34}z_{45}z_{56}z_{67}z_{73}}\,,\\
&\langle{\Psi_{(12)}\Psi_{(34)}\Psi_{(567)}}\rangle=-\frac{1}{12}\frac{1}{z_{12}^2z_{34}^2z_{56}z_{67}z_{75}}\nonumber\\*
&\qquad\qquad\times\Big[T_{12}\text{tr}(34)\text{tr}(567)+\text{tr}(12)T_{34}\text{tr}(567)\nonumber\\*
&\qquad\qquad\quad\;+\text{tr}(12)\text{tr}(34)T_{567}+T_{12}T_{34}\text{tr}(567)\nonumber\\*
&\qquad\qquad\quad\;+\text{tr}(12)T_{34}T_{567}+T_{12}\text{tr}(34)T_{567}\Big]\,.
\end{align}
\endgroup
Up to seven points, we find that $\langle\cdots\rangle$ simply averages over $\text{tr}(I)$ and $\tilde{T}_I$. We leave the study of generic cases to future works.

\bibliography{reference}

\end{document}